\documentclass[12pt]{article} 
\usepackage{amssymb}
\usepackage{graphicx}
\begin{document}  
\begin{center}
{\large \bf Unexpected properties of interactions of high energy protons}

\vspace{0.5cm}                   

{\bf I.M. Dremin$^{1,2}$}

\vspace{0.5cm}                       

        $^1$Lebedev Physics Institute, Moscow 119991, Russia\\
\medskip

    $^2$National Research Nuclear University "MEPhI", Moscow 115409, Russia     

\bigskip

{\large \bf Content}

\end{center}

{\bf 1.  Foreword

2. Introduction

3. Elastic scattering}

3.1. The diffraction cone 

3.2. The real part of the elastic scattering amplitude

3.3. The differential cross section outside the diffraction cone

{\bf 4. The unitarity condition

5. Central collisions

6. The shape of the inelastic interaction region at current energies

7. Some predictions at higher energies

8. Discussion and conclusions

References}

\bigskip

\begin{abstract}
Experimental data on proton-proton interactions in high energy collisions show 
quite a special and unexpected behaviour of the proportion of elastic scattering 
compared to inelastic processes with increasing energy. It decreases at the
beginning (at comparatively low energies) but then starts increasing. 
From Intersecting Storage Rings (ISR) energies of 23.5 - 62.5 GeV up to higher 
energies 7 - 13 TeV at the Large Hadron Collider (LHC) it increases by 
a factor more than 1.5! According to intuitive classical ideas we would expect 
a stable tendency with increasing proportion of the break-down of protons 
compared to their survival probability. One can assume
that either the asymptotic freedom or the extremely short time of flight of high
energy protons through each other are in charge of such a surprising effect. 
The unquestionable principle of unitarity combined with the available 
experimental data
on elastic scattering is used to get new conclusions about the shape of the
interaction region of colliding protons. Its evolution at present energies
is considered. Some predictions about its behaviour at even higher 
energies are described with different assumptions on relative roles of
elastic scattering and inelastic processes. The shape can transform rather
drastically if the proportion of elastic processes keeps rising. This unexpected
property leads to an unexpected corollary. The possible origin of the effect 
and its interrelation to the strong interaction dynamics are speculated.
\end{abstract}

\section{\it Foreword}

{\it If a cup falls to the floor, it breaks up to pieces but sometimes stays
intact. The harder it hits the floor, the less chance to be unbroken.

If two high energy protons collide, many new particles (mostly pions) are
produced, but sometimes they scatter elastically and retain their entity.
It is surprising enough that at very high collision energies the proportion of 
elastic processes increases with increasing energy from the ISR to the LHC. 

This unexpected and paradoxical phenomenon and its consequences at present 
and higher energies are discussed in the review.}

\section{Introduction}

One gets often accustomed to unexpected facts and they become just either the 
everyday reality or the trivial observation. However sometimes they stay 
unexplained for a long time.

In the 50-th, the strong interactions of hadrons impressed the physics community
by production of resonances in the pion-proton collisions. Afterwards, the
resonances filled in all the tables of elementary particles and became the
well known phenomenon. This process lasts up to now with the discovery of
the famous Higgs-boson or by the "closure" of the massive two-photon resonance.
The phemnomenon is described in terms of the dynamical levels of the system.

However not all the discoveries have the required interpretation.
In the beginning of the 70-th, it was unexpectedly found that the total cross 
section of the interaction of positively charged kaons with protons became 
increasing with the energy increase
already at energies of the Protvino accelerator up to 70 GeV in 
the rest (laboratory) system of one of the protons or about 12 GeV in the 
center-of-mass system. Let us remind that up to that time it was
commonly believed that hadronic cross sections must either decrease or tend
to constant values with energy increase. This belief was first strongly
shuttered by the so-called "Serpukhov effect". Nowadays it is well known that 
the total cross section of interaction of high energy protons steadily 
increases with the increase of energy of colliding partners. The elastic 
scattering cross section as well as the cross section of inelastic processes 
increase with energy also. Both the larger intensity of the interaction due to
the larger number of the actively participating partons (mostly, gluons)
and its larger spatial extention can be in charge of
that behaviour. Moreover, it happens that all hadronic cross sections increase 
with energy. Almost half a century has passed since then but no fundamental 
explanation of such behaviour in the quantum field theory has been proposed. 
Phenomenologically, it is usually described nowadays by the power-law
energy dependence due to exchange by the so-called supercritical Pomeron.
Its dynamical origin is yet unclear.
 
It is less known that experimental data hide another quite surprising and 
completely unexpected phenomenon of increase of the ratio of elastic to 
inelastic (or total) cross sections with energy increase in the interval 
from ISR (20-60 GeV in the center-of-mass system) \cite{ama, ame} to 
the highest explored accelerator energies at LHC (7-13 TeV) \cite{totem1, 
totem2, atlas}. The share of elastic collisions in the total outcome of all 
processes used to decrease at lower energies that coincided with our 
expectations. However, it reversed the tendency at ISR (the corresponding data 
were analyzed by me and the table with them was demonstrated earlier in 
Physics-Uspekhi journal \cite{ufnel, drem2}). Their relative 
roles evolve drastically. The inelastic cross section is about 5 times larger 
than the elastic one at ISR while their ratio decreases to 3 at LHC energies. 
According to the intuitive classical ideas we would expect the opposite 
behaviour with increasing probability of the break-down of both colliding 
protons into more and more "pionic pieces" compared to their survival 
probability, when protons are scattered purely elastically. Moreover, this 
increasing proportion of elastic scattering approaches such critical value at LHC 
energies \cite{jetp14, igse, ijmp} which, probably, indicates the transition 
to some principally new regime of interactions. Somehow the protons tend to
keep their entity while colliding with higher and higher energies.
No reliable explanation to this fact exists as well! Some simplest proposals
are considered only.

Here, we show the consequences of such an increase at present energies in the
picturesque presentation of the spatial interaction regions of colliding 
protons. We describe their possible non-trivial evolution 
at higher energies if this tendency persists. The adopted approach relies only 
on the unitarity condition and experimental data about elastic scattering of 
protons. No phenomenological input has been used. That assures the validity of 
conclusions. The results of some phenomenological models are discussed just to 
provide additional support to our statements. 

The general indubitable principle of conservation of total probability known 
in particle physics as the unitarity condition relates elastic and inelastic
processes. Sum of their ratios to total outcome should be equal 1. 
Therefrom, some knowledge about inelastic processes can also be 
gained using the elastic scattering data. The latter ones depend on smaller
number of variables. Thus they can be analyzed more easily. Surely, from
another side, that leads to the somewhat restricted sample of conclusions about
inelastic processes which one gets from the unitarity condition. Nevertheless,
one gains some knowledge about the spatial interaction region of protons at 
present energies and its possible evolution at higher energies. 

From the heuristic point of view, 
the increase of the share of elastic scattering to the critical value attained 
at LHC can for the first time reveal the transition from the traditionally 
considered branch of the unitarity condition dominated by inelastic processes 
(where elastic scattering is treated as the shadow of inelastic collisions) to 
another branch with the dominance of elastic scattering. That would require 
the completely new physical interpretation of the mechanism of proton (hadron) 
interactions and, probably, the formulation and further studies of new 
dynamical equations. 

The increase of the proportion of the elastic scattering processes reveals 
itself, first of all, in the spatial evolution of the elastic and 
inelastic interaction regions of colliding protons from ISR to LHC energies. 
It happened to be instructive to learn that the inelastic interaction region 
becomes more Black (absorptive) at the center, has steeper Edges (sharper 
decrease) and enLarges in size due to its periphery (the so-called BEL-scenario 
\cite{rhpv}) with energy increase in this energy interval. Even though the
form of these regions can not be measured directly in experiment, this 
knowledge has been used, for example, for interpretation of some peculiar 
features of experimental data on jet production at 7 TeV. Also, it
inspires theoretical ideas about possible 
experimental implications of their further evolution at higher energies. 
If the noticed tendency persists at higher energies, the profiles of both 
elastic and inelastic interactions can change drastically and show quite 
unexpected features, especially in the case of head-on collisions. Thus 
the BEL-scenario can be replaced by the absolutely new toroid-like regime
with the enlarged role of elastic scattering for central collisions.
It could be named as TEH-regime (Toroidal Elastic Hollow). 

No explanation of this phenomenon at present energies has yet been proposed.
What concerns our attempts to extrapolate it to higher energies, we hope that
experimental studies of elastic scattering of polarised protons or charge 
asymmetries of pions produced in inelastic collisions (or other yet
unexploited observations) could help in the proper choice of different 
possibilities. From the theoretical side, one can try to use more traditional 
QCD approach with enlarged fluctiations of gluon fields at collisions or 
revolutionary speculate on peculiar properties of solitons and instantons
using the corresponding equations in attempts to find a reasonable explanation.

Let us stress once again that he approximations adopted in the considered 
approach are completely justified so that one can claim that all results are 
obtained directly from combination of the two well-grounded sources - the 
unitarity condition and experimental data about elastic scattering and do not 
require any phenomenological input and modelling. Therefore the derived 
conclusions are very reliable. Their extrapolation to ever higher energy 
regions relies on the only assumption that the tendency of the increase
of the share of elastic scattering experimentally observed in the energy 
interval from ISR to LHC will persist there as well.

The structure of this review is as follows. In section 3 we start with the
description of general features of experimental results on elastic scattering 
of protons. Then the effective theoretical tool of the unitarity condition is 
introduced in section 4. There we discuss the accuracy of main approximations 
for elastic scattering amplitude which will be necessary for reliable 
estimates in the framework of the unitarity condition. It is applied further 
in section 5 to the special case of central head-on collisions of protons 
which allows to demonstrate typical features of unitarity constraints.
Then in section 6 the transverse spatial shapes of the inelastic and elastic 
interaction regions at current energies are demonstrated and their energy 
evolution is discussed. Possible extrapolations of the
profiles of interactions beyond modern (LHC) energies to asymptotics are 
presented in section 7 for different assumptions on the energy behaviour of
the proportion of elastic scattering. Finally, some conclusions are given
at the very end of the paper. Some assumptions about possible dynamical
origin of the observed effect are discussed as well.

\section{Elastic scattering}
 
The information about elastic scattering of protons comes from the measurement 
of the differential cross section $d\sigma /dt$ at some energy $s$ as a function 
of the transferred momentum $t$ at its experimentally accessible values.
It is related to the scattering amplitude $f(s,t)$ in the following way
\begin{equation}
\frac {d\sigma }{dt}=\vert f(s,t)\vert ^2 \equiv ({\rm Re}f(s,t))^2
+({\rm Im}f(s,t))^2.
\label{dsdt}
\end{equation}
The variables $s$ and $-t$ are the squared total energy $2E$ and the squared 
transferred momentum of the two colliding protons in the center-of-mass 
system $s=4E^2=4(p^2+m^2)$ ($p$ is the proton' momentum)
and $-t=2p^2(1-\cos \theta)$ at the scattering angle $\theta $. From this 
measurement one gets the knowledge only about the modulus of the amplitude,
i.e. about the sum of the squared values of its real and imaginary parts but
not about their signs. 
The Coulomb scattering contribution to it can be neglected everywhere except 
small angles. However, namely there the Coulomb scattering of the electrically
charged protons appears to be comparable to their nuclear interaction. The 
interference between the nuclear and Coulomb contributions to the amplitude $f$ 
becomes quite large and allows to find out from the shape of the experimental 
differential cross section the ratio of the real and imaginary parts of the 
elastic scattering amplitude $\rho (s,t)={\rm Re}f(s,t)/{\rm Im}f(s,t)$.
This can be done just in forward direction $t=0$ $\rho (s,0)=\rho _0$ 
(to be more precise, extremely close to it) but not at any other values of $t$.

The typical shape of the experimentally measured differential cross section 
at high energies shown in Figures contains some characteristic features.
Those are the above mentioned interference region at extremely small values of 
$\vert t\vert $, almost invisible in Fig. 1,
the exponentially decreasing (with increase of $\vert t\vert $) diffraction 
cone with energy dependent slope $B(s)$ (Fig. 1), the dip (Fig. 2) and more 
slowly decreasing tail at larger transferred momenta with much smaller values 
of the cross section compared to the diffraction cone (Fig. 2).

\begin{figure}
\includegraphics[width=\textwidth, height=8cm]{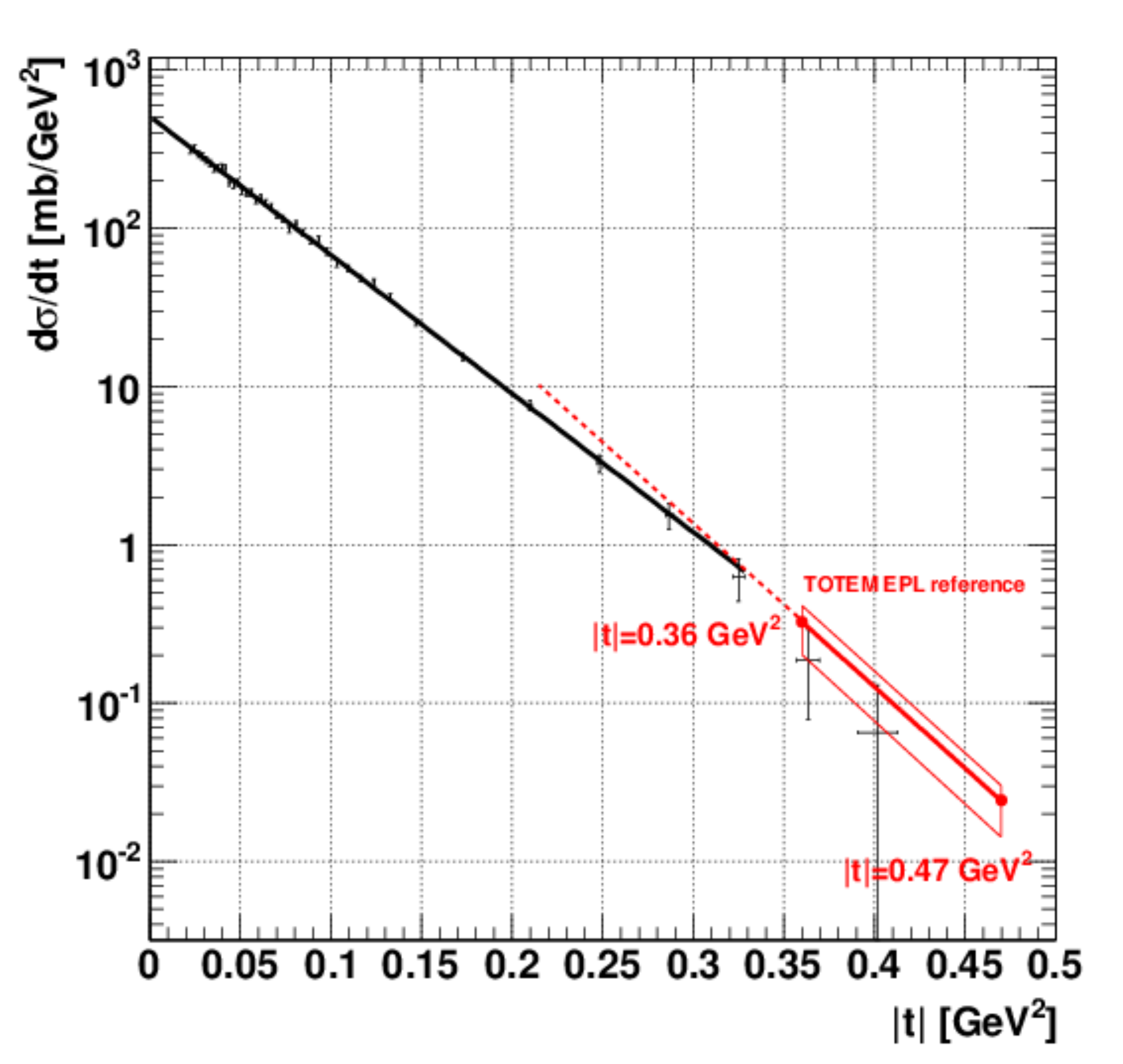}

Fig. 1. The differential cross section of elastic proton-proton scattering at 
$\sqrt s$=7 TeV measured by the TOTEM collaboration  
(Fig. 4 in \cite{totem1}). \\
The region of the diffraction cone with the $\vert t\vert $-exponential 
decrease is shown.

\end{figure}

\begin{figure}
\centerline{\includegraphics[width=\textwidth]{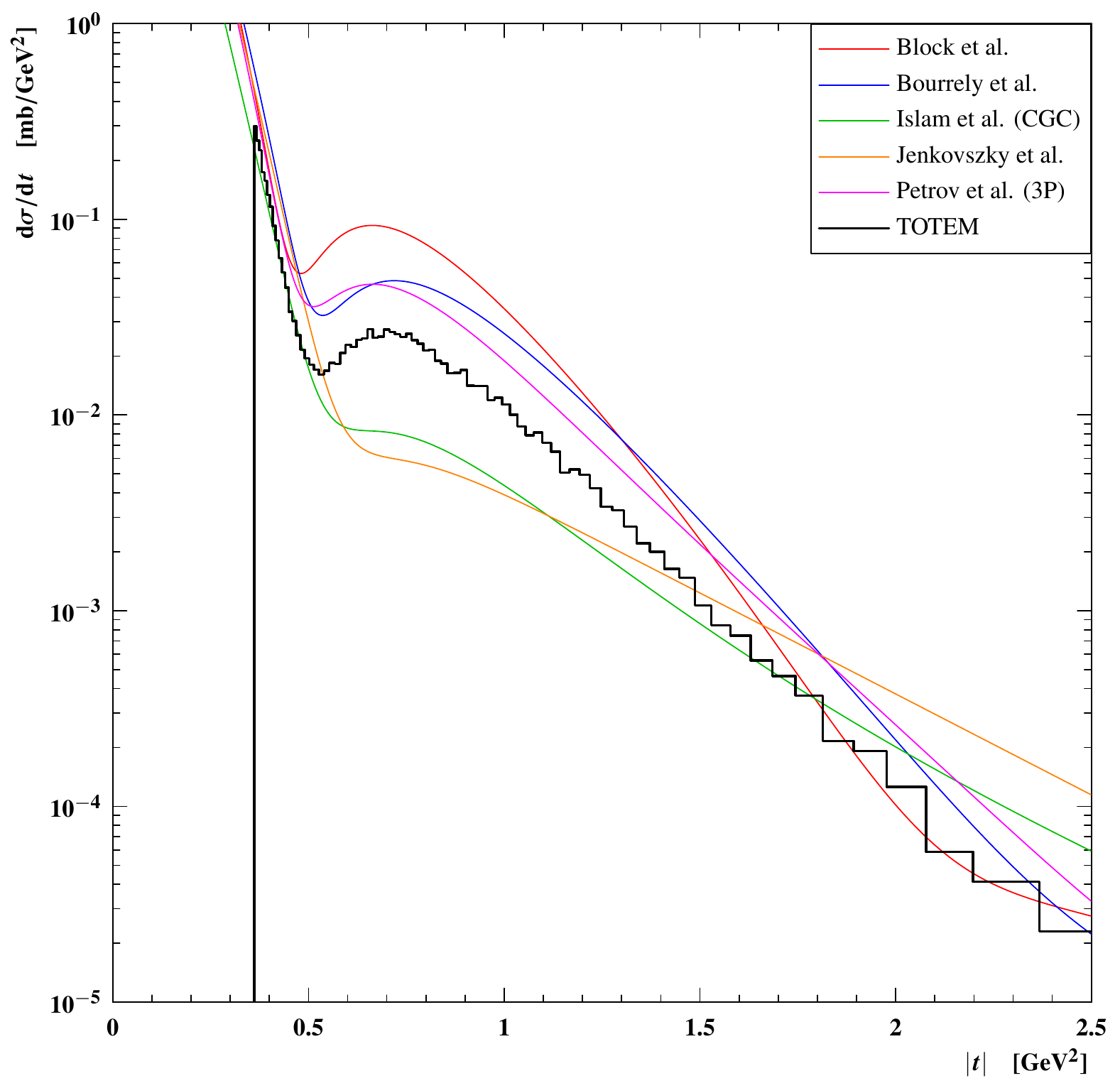}}

Fig. 2. The differential cross section of elastic proton-proton scattering at 
$\sqrt s$=7 TeV measured by the TOTEM collaboration  
(Fig. 4 in \cite{totem2}). \\
The region beyond the diffraction peak is shown. The predictions of five 
phenomenological models are demonstrated. 
\end{figure}

\subsection{The diffraction cone}

The diffraction cone is shown in Fig. 1. Protons scatter mainly at processes 
with small transferred momenta. The differential cross section is much 
larger there than at larger transferred momenta. Its exponential 
parameterization is demonstrated by the straight line at the logarithmic scale. 

There are several tiny features of this plot. In the very narrow region of 
extremely small transferred momenta the 
amplitude is represented by the sum of the nuclear and Coulomb amplitudes.
Their interference produces some increase of the differential cross section 
in there. It has been used for estimates of the real part of the amplitude. 
Moreover small deviations of the order of 1 per cent from the exponential
shape (invisible in Fig. 1) were noticed at the extremely precise measurements 
at 8 TeV \cite{totem2}. Also one can see the somewhat steepened shape at the 
very end of the diffraction cone approximated there by another exponent 
(the dashed line) which differs from the leading one albeit not very strongly
and the whole effect is noticeable only in a very small interval of transferred
momenta. Let us note that at lower energies the 
shape was slightly flattened but not steepend. The impact of all these 
specific features on our further calculations is easily estimated. It will
be shown very small bcause we will use the averaged integrated parameters.
Threfore in what follows we adopt the simple exponential parameterization of 
the diffraction cone which is precise enough for the corresponding transferred
momenta and has been used by experimentalists:
\begin{equation}
\vert f(s,t)\vert \approx \frac {\sigma _{tot}(s)}{4\sqrt {\pi}} \exp [B(s)t/2]
\label{fst}
\end{equation}
where $\sigma _{tot}(s) $ is the total cross section and $B(s)$ is the energy
dependent slope of the diffraction cone.

\subsection{The real part of the elastic scattering amplitude}

Some theoretical information about the energy behaviour of the real part of the 
forward scattering amplitude can be obtained from the dispersion relations 
which follow from the analyticity property of the amplitude. They 
relate it to the integral of the imaginary part at zero angle, i.e. 
to the total cross section according to the optical theorem (see Eq.
(\ref{opt}) below). Using reasonable extrapolations of the total cross 
section to higher energies it was predicted long ago \cite{dnaz, blo, bloh}
that at high energies the real part is small compared to the imaginary part and
their ratio is about 0.12 - 0.15 with slow decrease at asymptotic energies.
Both real and imaginary parts are positive at $t=0$ due to positivity of the
latter. These predictions were confirmed by experiment. At LHC energies the 
measured ratios range is 0.12 - 0.145 \cite{totem1, atlas, totem4}. Thus the 
real part only contributes about 1 - 2$\%$ to the differential cross section 
(\ref{dsdt}) at $t=0$. 

What concerns the behaviour of the real part as a function of the transferred 
momentum, some general theoretical guesses \cite{mart1, mart2} indicated 
that it can become zero somewhere within the diffraction cone. Therefore
its decrease inside the diffraction cone should be steeper than for the
imaginary part, and, consequently, its integral contribution from this region 
to the elastic cross section must be even smmaller. No definite 
position was ascribed in the papers \cite{mart1, mart2} to the point 
where it crosses the abscissa axis. Recently, some possibilities to use
the analytical properties of the elastic scattering amplitude for
getting some knowledge about its real part were considered in Ref. \cite{ann}.

Nevertheless, one can easily estimate from the data presented in Fig. 1 and
Fig. 2 the upper limit of the real part of the amplitude at the dip. Its ratio
to the imaginary part at $t=0$ is calculated as the square root of the ratio
of the differential cross sections at those points and, surely, is very
small $\leq 0.006$. This estimate supports our intention to neglect the 
contribution of the real part of the amplitude in further calculations
where its integrally averaged characteristics are only used.

Further guides about its behaviour can only be obtained from particular models 
of proton interactions. Those of them which pretend to make precise fits of 
a wide variety of present experimen tal data are surely preferred. Even then
they should not be absolutely trusted because we have some experience that 
several details got wrong even at present energies and could become worse
at extrapolations to new energy fields. Nevertheless, 
as such an example, we show in Fig. 3 borrowed from Ref. \cite{kfk} the 
behaviour of the real and imaginary parts of the elastic scattering amplitude 
at energy 7 TeV within the large interval of the transferred momenta. 
Its shape is derived with the help of a particular  phenomenological
model \cite{kfk} which happened to be very successful in fits of many 
experimental characteristics in a wide range of energies up to
LHC. 

In particular, one can see that the real part at 7 TeV is much 
 smaller than the imaginary part everywhere within the diffraction cone
and crosses the abscissa axis in accordance with theoretical expectations
\cite{mart1, mart2}. Its relative contribution to the differential cross 
section (\ref{dsdt}) is given by the term $\rho ^2(s,t)$ where 
$\rho (s,t)={\rm Re}f(s,t)/{\rm Im}f(s,t)$. It can be neglected in the
model considered. The accuracy of experimental data is not yet high enough
for such small contributions to be taken into account. That corresponds well 
with our prejudice that the diffraction cone is somehow a shadow of inelastic 
processes because the elastic amplitude is substantially imaginary there. It is
interesting to note that according to the model \cite{kfk} the imaginary part 
dominates everywhere besides the dip interval which is very short. However, the 
differential cross section is already very small there compared to the
diffraction cone. Thus in our analytical estimates we will neglect the real 
part of the amplitude but sometimes come back to it to show once again how
irrelevant for our conclusions is its contribution.

\begin{figure}[hbtp]
\begin{center}
\includegraphics[width=\textwidth, height=7.4cm]{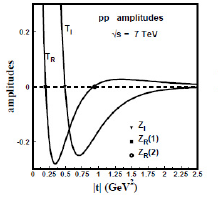}%

Fig. 3. Real ${\rm Re}f$ and imaginary ${\rm Im}f$ parts of the proton-proton amplitude 
at 7 TeV according to a particular phenomenological model \cite{kfk}. Note that 
the contribution of the real part to $d\sigma /dt$ becomes noticeable only
near the dip where $d\sigma /dt$ is small. It can be completely neglected 
inside the diffraction cone. Moreover, it becomes equal zero inside it as was
predicted \cite{mart1, mart2}. It is quite interesting that the imaginary part 
dominates in the Orear region of intermediate transferred momenta as well.
\end{center}
\end{figure}
The steep exponential decrease of differential cross sections in the diffraction 
cone implies that namely this region contributes mostly to Eq. (\ref{gamm}.
The integral contribution of the real part of the amplitude $f$ in there must 
even be noticeably smaller than its overestimated value $\rho _0{\rm Im}f$. 
That is why it is possible to neglect it further in analytical calculations.

\subsection{The differential cross section outside the diffraction cone}

In what follows we will need to estimate the contribution of the region outside 
the diffraction peak to some analyzed variables. Comparison of Fig. 2
with Fig. 1 shows that the differential cross section is much lower (by more
than 4 orders of magnitude!) at the dip and at the tail compared to its values
at the beginning of the diffraction cone. Moreover, it decreases approximately 
as $\exp (-c(s)\sqrt {\vert t\vert })$ in this region. It is usually called as
the Orear region by the name of its first observer and can be explained (see 
\cite{anddre1}) by subsequent iterations (rescattering) in the solution of 
the unitarity equation in the ($s,t$)-representation. Surely, the 
$t$-exponential parameterization (\ref{fst}) used for the diffraction cone
underestimates the contribution of the tail with 
$-\sqrt {\vert t\vert }$-exponent at high transferred momenta. 
However, the integral contribution to our variables of the excess in the tail 
region over our approximation (\ref{fst} is easily estimated. We show below 
that it is negligibly small. The interplay of the real and imaginary 
parts of the amplitude $f$ can be more complicated there as seen, e.g., from 
Fig. 3. However, the smallness of the modulus, i.e., of $\sqrt {d\sigma /dt}$, 
implies the smallness of both of them in this region even though their ratio
$\rho $ becomes infinitely large if the imaginary part becomes equal to zero. 

\section{The unitarity condition}

Our main goal here is to get some knowledge about the spatial region of 
interactions of high energy protons at current energies, to draw a pictorial 
view of its evolution with increasing energy and to discuss possible 
theoretical and experimental implications of these findings.

The most stringent and reliable information (albeit rather limited!)
about the interrelation of elastic and inelastic processes comes 
from the unitarity of the $S$-matrix 
\begin{equation}
SS^+=1
\label{ss}
\end{equation}
or for the scattering matrix $T \; (S=1+iT)$
\begin{equation}
2{\rm Im}T_{ab}=\Sigma _n\int T_{an}T^*_{nb}d\Phi _n,
\label{tt}
\end{equation}
where $a, b, n$ denote the number of particles. The whole $n$-particle phase 
space $\Phi _n$ is integrated over. For the elastic scattering amplitude 
$a=b=2$, the unitarity condition 
relates the amplitude of elastic scattering $f\propto T_{22}$ to the amplitudes 
of $n$-particle inelastic processes $T_{2n}$ declaring that the total 
probability of all outcomes of the interaction (elastic and inelastic ones)
must be equal to 1\footnote{The non-linear contribution from the elastic 
amplitude appears in the right-hand side for $n=2$.}.

In the $s$-channel this uinquestionable condition is usually expressed in the form 
of the well known integral relation (for more details see, e.g., 
\cite{PDG, anddre1, ufnel}). This relation is quite complicated for arbitrary 
values of the transferred momentum, $t$. However, for forward scattering at 
$t=0$ it leads to the widely used optical theorem showing the normalization of 
the imaginary part of the amplitude ${\rm Im}f(s)$ 
by its direct connection with the total cross section $\sigma _{tot}$:
\begin{equation}
{\rm Im}f(s,0)=\sigma _{tot}(s)/4\sqrt {\pi}
\label{opt}
\end{equation}
and to the general statement that the total cross section is the sum of 
cross sections of elastic and inelastic processes
\begin{equation}
\sigma _{tot}=\sigma _{el}+\sigma _{inel},
\label{telin}
\end{equation}
i.e., that the total probability of all processes equals 1.

One can use the Fourier -- Bessel transform of the amplitude $f$ to reduce
the integral relation to the more simple algebraic one. This transformation
retranslates the momentum data to the shortest transverse distance between 
the trajectories of the centers of colliding protons called the impact 
parameter, $b$, and is written as                             
\begin{equation}
i\Gamma (s,b)=\frac {1}{2\sqrt {\pi }}\int _0^{\infty}d\vert t\vert f(s,t)
J_0(b\sqrt {\vert t\vert }).
\label{gamm}
\end{equation}
Then the unitarity condition in the $b$-representation reads 
\begin{equation}
G(s,b)=2{\rm Re}\Gamma (s,b)-\vert \Gamma (s,b)\vert ^2.
\label{unit1}
\end{equation}
(for reviews see, e.g., Refs \cite{ufnel, drem2}).
This relation establishes the connection between the distributions of the
intensity of all processes in the transverse configuration space:
\begin{equation}
\frac {d^2\sigma _{inel}}{db^2}=\frac {d^2\sigma _{tot}}{db^2}-
\frac {d^2\sigma _{el}}{db^2}
\label{elin}
\end{equation}
The left-hand side in Eqs (\ref{unit1}), (\ref{elin}) describes the transverse 
impact-parameter profile of inelastic collisions of protons. It satisfies 
the inequalities $0\leq G(s,b)\leq 1$ and determines how absorptive is the 
interaction region at the given impact parameter (with $G=1$ for the full 
absorption and $G=0$ for the complete dominance of elastic scattering). 
The profile of elastic 
processes is determined by the subtrahend in Eqs (\ref{unit1}),
(\ref{elin}). Thus we get a spatial view of the whole process if the elastic 
scattering amplitude $f$ is integrated in Eq. (\ref{gamm}).

Let us note from the very beginning that these profiles can not be measured 
directly in experiments because the impact parameters are not the measurable
quantities. Nevertheless, their energy behaviour has important heuristic value
because it can reveal the evolution of the process dynamics. It will be 
described below how the knowledge of the spatial extension of the
inelastic interaction region has been used for the description of the
processes of jet production at the LHC energy 7 TeV. One can use
various models of the interactions and confront different assumptions.
Also, one can try to relate the impact parameters, for example, with the
multiplicities of inelastic collisions as it is done for the interactions
of the relativistic nuclei. However, we do not speculate on it in this review. 

If $G(s,b)$ is integrated over the impact parameter, it leads to the cross 
section of inelastic processes. The terms on the right-hand side of Eqs 
(\ref{unit1}), (\ref{elin}) would correspondingly produce the total cross 
section and the elastic cross section in accordance with Eq. (\ref{telin}).

It follows from the above relations that, strictly speaking, one should know 
both real and imaginary parts of the elastic scattering amplitude to get results 
about the impact-parameter profiles of inelastic and elastic processes from the 
unitarity condition. However, its modulus can only be found from experimental 
data as follows from Eq. (\ref{dsdt}) and some very limited knowledge 
about its real part for forward scattering. Nevertheless, one can easily
estimate the accuracy of any assumption in calculations according to 
Eqs. (\ref{gamm}), (\ref{unit1}).

In particular, we will use the fact that the modulus of the amplitude decreases 
approximately exponentially (see Eq. (\ref{fst})) in the diffraction cone 
and becomes much smaller at the tail compared to its values at the top of the 
diffraction peak. The slight decline from a simple exponent inside the cone of 
the order of 1$\%$ noticed recently by TOTEM Collaboration \cite{totem3} at small
transferred momenta as well as somewhat steepened behaviour at the very end of 
the diffraction cone near the dip seen in Fig. 2 do not influence its integral 
contribution to (\ref{gamm}) within the accuracy of determination of the slopes. 
In what follows, we use the exponential parametrization of the imaginary 
part of the amplitude $f$ to proceed with analytical calculations and argue
that it is very precise: 
\begin{equation}
{\rm Im}f(s,t)=\frac {\sigma _{tot}(s)}{4\sqrt {\pi}} \exp [B(s)t/2].
\label{opt1}
\end{equation}
Formally, this approximation is not valid for differential cross sections at 
large transferred momenta. However, for our purposes the integral
contribution of $f$ at large $\vert t\vert $ to Eq. (\ref{gamm}) is only 
important. It is negligibly small there compared to the peak of the diffraction 
cone. The approximation (\ref{opt1}) is justified as will be shown below. In 
fact, that was clear earlier when it was demonstrated \cite{dnec} that such 
approximation and direct integration of experimental data lead to the practically 
indistinguishable results. The accuracy of calculations is very high. Thus one 
can claim that the resuts obtained analytically rely only on the unitarity 
condition and experimentally measured exponential decrease of the differential
cross section in the diffraction cone. 

\section{Central collisions}

Before using the detailed formulae for the spatial extension of the interaction
region as a function of the impact parameter $b$, let us study at the beginning 
the simpler case of the energy dependence of the intensity of interaction for 
central (head-on) collisions of impinging protons at $b=0$. We introduce the
variable $\zeta $:
\begin{equation}
\zeta (s)={\rm Re}\Gamma (s,0).
\label{zeta1}
\end{equation}
For the dominant contribution of the diffraction cone (Eq. (\ref{opt1})) one
gets that $\zeta $ is directly related to the share of elastic processes:
\begin{equation}
\zeta (s)=\frac {4\sigma_{el}}{\sigma _{tot}}
\label{zeta}
\end{equation}
One can also write
\begin{equation}
\vert \Gamma \vert ^2=\zeta ^2+\frac {1}{4\pi}\left(\int _0^{\infty}d\vert t\vert 
{\rm Re}f\right) ^2.
\label{g2}       
\end{equation}
The last term here can be neglected compared to the first one. That is easily
seen from
\begin{equation}
\int _0^{\infty}d\vert t\vert {\rm Re}f \leq
\int _0^{\infty}d\vert t\vert \vert {\rm Re}f\vert =
\int _0^{\infty}d\vert t\vert \sqrt 
{\frac {\rho ^2(s,t)d\sigma /dt}{1+\rho ^2(s,t)}}.
\label{g3}       
\end{equation}
The factor $\rho ^2(s,t)/(1+\rho ^2(s,t))$ is very small in the diffraction 
cone. It can become of the order 1 at large values of $\rho ^2(s,t)$ 
(say, at the dip) but the cross section is small there already (compare Fig. 2 
and Fig. 3). Then the unitarity condition (\ref{unit1}) is written as 
\begin{equation}
G(s,b=0)= \zeta (s) (2-\zeta (s)).
\label{gZ}       
\end{equation}
Thus, according to the unitarity condition (\ref{gZ}) the darkness of the 
inelastic interaction region for central collisions (absorption) is defined 
by the only experimentally measured parameter $\zeta (s)$ depending on energy. 
It has the maximum $G(s,0)=1$ for $\zeta =1$. Any decline of $\zeta $ from 1
($\zeta =1\pm \epsilon $) results in the parabolic decrease of the absorption
($G(s,0)=1-\epsilon ^2$), i.e. to an even much smaller decline from 1 for
small $\epsilon $. The elastic profile, equal to $\zeta ^2$ in central 
collisions, also reaches the value 1 for $\zeta =1$.

The unitarity condition imposes the limit $\zeta \leq 2$ on the increase of 
the share of elastic scattering. It is required by the positivity of the 
inelastic profile. Then there no inelastic processes for central collisions 
($G(s,0)=0$ according to Eq. (\ref{gZ})). This limit corresponds to the widely
discussed "black disk" picture which asks for the relation
\begin{equation}
 \sigma _{el}=\sigma _{inel}=\sigma _{tot}/2.
\end{equation}
The value of the profile of central ($b=0$) elastic collisions $\zeta ^2$ 
completely saturates the total profile $2\zeta $ for $\zeta =2$. Below, 
we shall discuss physics implications of these findings.

With high enough precision one can describe $\zeta $ by the following 
formulae:
\begin{equation}
\zeta (s)\approx \frac {\sigma _{tot}(s)}{4\pi B(s)}\approx (4\pi )^{-0.5}
\int _0^{\infty }d\vert t\vert \sqrt {\frac{d\sigma /dt}{1+\rho ^2(s,t)}}.
\label{ze}
\end{equation}
One should specially note that all formulae contain only experimentally
measurable quantities $\sigma _{tot}(s), \sigma _{el}(s), B(s)$. 
The most convenient for our further discussion is its interpretation as a 
share of elastic processes because, in particular, it is proportional to the 
experimentally measurable dimensionless ratio of the elastic cross section 
$\sigma _{el}$ to the total cross section $\sigma _{tot}$ (\ref{zeta}).

From the first formula one gets the conclusion that the increase of $\zeta (s)$
with increasing energy demonstrates that the height of the diffraction cone
(the numerator) increases faster than its width shrinks (the denominator).

From the second relation in (\ref{ze}) one can get very definite conclusions 
about the role of different regions of the differential cross section for the 
variable $\zeta $ and, consequently, for the unitarity condition. In practice, 
one should just integrate the squared root of the differential cross section
over the corresponding interval of transferred momenta. It is clearly seen
that its value is mainly determined by such transferred momenta where
the differential cross section is large and the real part of the amplitude 
is small compared to the imaginary part. This is valid in the diffraction cone.
The simplest estimates with constant value $\rho _0(s)\approx 0.02$ in place 
of $\rho (s,t)$ in Eq. (\ref{ze}) show that this contribution is at the level
of 1$\%$. It is greatly reduced if its values from Fig. 3 are used since
the values of ${\rm Re}f $ are smaller there and, moreover, their contribution
is exponentially weighted within the diffraction cone in (\ref{ze}).
Surely, one can neglect by small declines from the simple exponential
shape both inside and at the end of the diffraction cone because their
contribution becomes very small after integration in Eq. (\ref{ze}). 
In fact, one can definitely state that the exponential 
parameterization of the imaginary part of the amplitude (\ref{opt1})
can be used for description of experimental data in our formulae.
The conclusions of the phenomenological model \cite{kfk} just support our
estimates as shown in \cite{dnw}. 

What concerns the tail of the differential cross section, the convenient 
approximation of $d\sigma /dt$ by a pure exponential (in Eq. (\ref{opt1})) 
is most easily verified by taking directly the published distribution and
carrying out the integration directly using the measured data. Numerically we 
find that the data, when the region above the dip are included, yield values 
of $\zeta $ which are less than 3.9$\% $ higher than obtained with the 
exponential approximation.  

This results in a less than $2\cdot 10^{-3}$ correction to the 
calculation of $G(7 \,$TeV,0) in Eq. (\ref{gZ}). These 2 approximations 
($\rho_0^2\approx 0.02$ and exponential form) allow us to greatly simplify 
the discussion of the profile function, and are, in any case, not contradicted 
by known data and experimental uncertainties. The discussion 
of the accuracy of estimates can be found in \cite{white}. 

The more detailed estimates of different contributions according to the
phenomenological model \cite{kfk} are given in Ref. \cite{dnw}. The 
imaginary part of the amplitude becomes negative after the dip in this
model. The contribution to the definition of $\zeta $ is also negative.
Its numerical value becomes lowered but again within several percents only.

The experimentally measured proportion of elastic processes 
$\sigma _{el}/\sigma _t=0.25 \zeta $ 
demonstrates the non-trivial dependence on energy shown in the Table. 
The values of the absorption at central collisions $G(s,0)$ and the ratios
of inelastic to elastic cross sections $\sigma _{in}/\sigma _{el}$  are 
also shown. All values are derived directly from experimental data at
corresponding energies $s$.
\medskip
\begin{table}
\medskip
Table.  $\;$ The energy behaviour of $\zeta $, $G(s,0)$ and 
$\sigma _{in}/\sigma _{el}$.
\medskip

    \begin{tabular}{|l|l|l|l|l|l|l|l|l|l|l|l}
        \hline
$\sqrt s$, GeV&4.11&4.74&7.62&13.8&62.5&546&1800&7000\\ \hline
$\zeta $           &0.98&0.92&0.75&0.69&0.67&0.83&0.93&1.00-1.04 \\  
$G(s,0)$     &1.00&0.993&0.94&0.904&0.89&0.97&0.995&1.00 \\  \hline
             & & & & & ISR&S$p\bar p$S &FNAL & LHC\\   \hline
$\sigma _{in}/\sigma _{el}$ & & & & &$\bf 5$ & & &$\bf 3$ \\  \hline
\end{tabular}
\end{table}
The change of the tendency in the behaviour of elastic processes with energy 
increase looks especially surprising. One would naively expect that their 
proportion would decrease being replaced by inelastic processes with higher 
multiplicities at higher energies. That happens only at low energies up to ISR 
where the parameter $\zeta $ decreases from about 1 down 
to values about 2/3. At higher energies protons reveal unexpected stability.
The share of elastic scattering increases with energy. The parameter $\zeta $ 
reaches the critical value 1 for 7 TeV data at LHC where the elastic cross 
section is about 4 times less than the total cross section. 

That looks even more impressive in terms of the ratio of the inelastic cross
section to the elastic one
\begin{equation}
\frac {\sigma _{inel}}{\sigma _{el}}=\frac {4}{\zeta }-1
\label{inel}
\end{equation}
The ratio decreases from 5 at ISR to 3 at LHC as shown in the Table.

It is intriguing whether this increase of the proportion of elastic scattering 
will really show up in experiments at higher energies or it will be saturated 
asymptotically with $\zeta $ tending to 1 from below. The asymptotic 
saturation would lead to the conservative stable situation on the same branch
of the unitarity condition while further increase above 1 will require 
the transition to another branch of the unitarity equation and new physics 
interpretation.
 
To explain the last statement let us rewrite Eq. (\ref{gZ}) as
\begin{equation} 
 \zeta (s)=1\pm \sqrt {1-G(s,0)}.
\label{zeta2}
\end{equation}
The critical value $\zeta =1$ reveals itself in the usage of different signs
in front of the square root term (different branches of the unitarity
condition) for $\zeta <1$ and $\zeta >1$.
One used to treat elastic scattering as a shadow of inelastic processes. This 
statement is valid when the branch with negative sign in Eq. (\ref{zeta2})
is considered because it leads to proportionality of elastic and inelastic 
contributions ($\zeta \propto G(s,0)/2$) for small $G(s,0)\ll 1$. That 
is typical for electrodynamical forces in particle interactions (e,g., for 
processes like $ee\rightarrow ee\gamma$) and for optics (photon interactions)
where the inelastic cross sections are small and their values are governed
by the fine structure constant $\alpha $. The large value of the inelastic
cross sections in hadronic collisions with subsequent increase of the elastic 
proportion at diminishing role of inelastic production destroys the analogy. 
That is why the observation of this effect comes as a surprise. For strong 
interactions, the shares of inelastic and elastic processes are compatible 
(see the Table). The approach 
of $\zeta $ to 1 at 7 TeV corresponds to complete absorption in central 
collisions. This value is considered as a critical one because from (\ref{zeta2}) 
one gets significant conclusion that the excess of $\zeta $ over 1 implies that
the unitary branch with positive sign in front of inelastic processes
is at work. This branch was first considered in \cite{trt} with application 
to high energy particle scattering. That changes the interpretation of the role 
of elastic processes as being a simple "shadow" of inelastic ones.

Present experimental data at LHC can not distinguish definitely between the two 
possibilities of asymptotic saturation and increase of the elastic share. 
Some slight trend of $\zeta $ to increase and become larger than 1 can be 
noticed from comparison of TOTEM data at 7 TeV \cite{totem1} where it can 
be estimated\footnote{The experimental values of the ratios of elastic to total 
cross section and $\rho _0$ have been used.} in the limits 1.00 and 1.04 and 
at 8 TeV \cite{totem2} where according to the data of the same collaboration
it is approximately 1.05 though within the accuracy of experimental data about 
$\pm $0.024. The data of ATLAS collaboration at 8 TeV do not reveal any increase 
of the proportion of elastic scattering albeit with approximately the same
accuracy. The more precise data at these energies and at 13 TeV are needed.

The further increase of the share of elastic scattering with energy is favored 
by extensive fits of available experimental information for the wide energy 
range and their extrapolations to ever higher energies done in 
the phenomenological models of Refs 
\cite{kfk, fms} as well as by some theoretical speculations (e.g., see Ref. 
\cite{roy}). The asymptotical values of $\zeta $ are about 1.5 in Refs
\cite{kfk, fms} and 1.8 \cite{roy}. They correspond to incomplete but rather
noticeable decrease of the absorption at the center of the interaction region. 
The corresponding values of the attenuation at the center $G(\infty ,0)$ are
0.75 and 0.36. It is discussed in more detail in the next Section.

\section{The shape of the inelastic interaction region at current energies}

The detailed shape of the inelastic interaction region at arbitrary values
of the impact parameters can be obtained with 
the help of relations (\ref{gamm}), (\ref{unit1}) if the behaviour of the 
amplitude $f(s,t)$ is known. Its modulus and the $\rho _0$ values are
obtained from experiment. The most prominent feature of experimental results 
at present energies from ISR to LHC is the rapid exponential decrease of 
$d\sigma /dt $ with increasing transferred momentum $\vert t\vert $, 
especially in the near forward diffraction cone. 
It is just this region of transferred momenta which contributes mostly to
Eqs (\ref{ze}), (\ref{gamm}).
Inserting the exponential shape of the cone in there one can write
\begin{equation}
i\Gamma (s,b)\approx \frac {\sigma _{tot}(s)}{8\pi }\int _0^{\infty}
d\vert t\vert \exp (-B(s)\vert t\vert /2 )(i+\rho (s,t))J_0(b\sqrt {\vert t\vert }).
\label{gam2}
\end{equation}
Let us stress that the diffraction cone dominates the contribution to 
${\rm Re}\Gamma $ in Eqs (\ref{zeta}), (\ref{gam2}) so strongly that the tail of 
the differential cross section at larger $\vert t\vert $ can be completely 
neglected at the level of some per cents by itself even for central collisions 
as was estimated in the previous chapter. Besides, it is suppressed 
additionally by the Bessel function $J_0$ at larger impact parameters. 
Therefore the accuracy of the approximation increases. It was 
estimated using fits of the experimental differential 
cross section outside the diffraction cone by simplest analytical expressions.
Moreover, it was shown \cite{dnec, ads} by computing how well the versions 
with direct fits of experimental data and with their exponential approximation 
coincide if used in the unitarity condition.  Therefore the expression 
(\ref{opt1}) can be treated as following directly from experiment and
being very precise. Herefrom, one calculates
\begin{equation}
{\rm Re}\Gamma (s,b)= {\zeta }{\exp (-\frac {b^2}{2B})}.
\label{rega}
\end{equation}
Correspondingly, the shape of the inelastic profile for small $\rho _0$ is 
given by
\begin{equation}
G(s,b)=  \zeta \exp (-\frac {b^2}{2B})[2-\zeta \exp (-\frac {b^2}{2B})].
\label{ge}
\end{equation}
It scales as a function of $b/\sqrt {2B}$. Its energy dependence is determined
by the two measured quantities - the diffraction cone width $B(s)$ and its 
ratio to the total cross section, i.e. by the variable $\zeta (s)$. 
It has the maximum at
\begin{equation}
b_m^2=2B\ln \zeta
\label{bm}
\end{equation}
It is positioned in the unphysical region of impact parameters $b_m^2<0$ for 
$\zeta <1$, i.e. at all energies below LHC. Therefore the absorption is 
incomplete $G(s,b)<1$ at any physical impact parameter $b\geq 0$. Its largest 
value is reached at the very center $b=0$. The inelastic interaction region 
has the shape of a disk with absorption strongly diminishing to its edges.
The disk is semi-transparent at ISR energies. This is demonstrated by
the corresponding line ($\zeta =0.7$) in Fig. 4 \cite{igse} shown below.

At $\zeta =1$, which is only reached at LHC energy 7 TeV, the maximum is 
positioned exactly at the center $b=0$ and the maximum absorption occurs just 
for central collisions, i.e. $G(s,0)=1$. The disk center becomes black. 
The strongly absorptive core of the inelastic interaction region 
grows in size compared to ISR energies (see \cite{dnec}) because of increase
of the slope $B(s)$. The enlarged size of the inelastic interaction region
can be clearly seen from the Taylor expansion of Eq. (\ref{ge}) at small 
impact parameters:
\begin{equation}
G(s,b)= \zeta [2-\zeta -\frac {b^2}{B}(1-\zeta )-\frac {b^4}{4B^2}(2\zeta -1)].
\label{gb}
\end{equation}
The negative term proportional to $b^2$ vanishes at $\zeta =1$, and $G(b)$ 
develops a wide strongly absorbing plateau which extends to the 
comparatively large values of impact 
parameters $b$ (up to about 0.5 fm). The plateau is very flat because the last 
negative term in Eq. (\ref{gb}) which diminishes the absorption starts to play 
a role at 7 TeV (where $B\approx 20$ GeV$^{-2}$) only for larger values of $b$.
Therefore the absorption decrease becomes steeper at the periphery. The earlier 
proposed scenarium BEL is therefore realized at present energies in such a way.
The two lines in Fig. 4 demonstrate the evolution of the shape of the inelastic 
interaction region from ISR ($\zeta =0.7$) to LHC ($\zeta =1.0$) energies. 
The larger darkness for central collisions at LHC compared to ISR can be 
ascribed to the enlarged role of soft gluons in the proton structure function.
It is claimed in several papers \cite{mart, also, arbr} that already at the LHC
energies the hollowness of the plateau can be seen at $b=0$. Actually,
the accuracy of experiments there is still not enough for the definite
conclusions. Only at higher energies (or if higher accuracy at LHC would be
achieved) it can be definitely observed as displayed in Fig. 4. We discuss
these predictions in the next section. 

\begin{figure}
\begin{center}
\includegraphics[width=\textwidth, height=9cm]{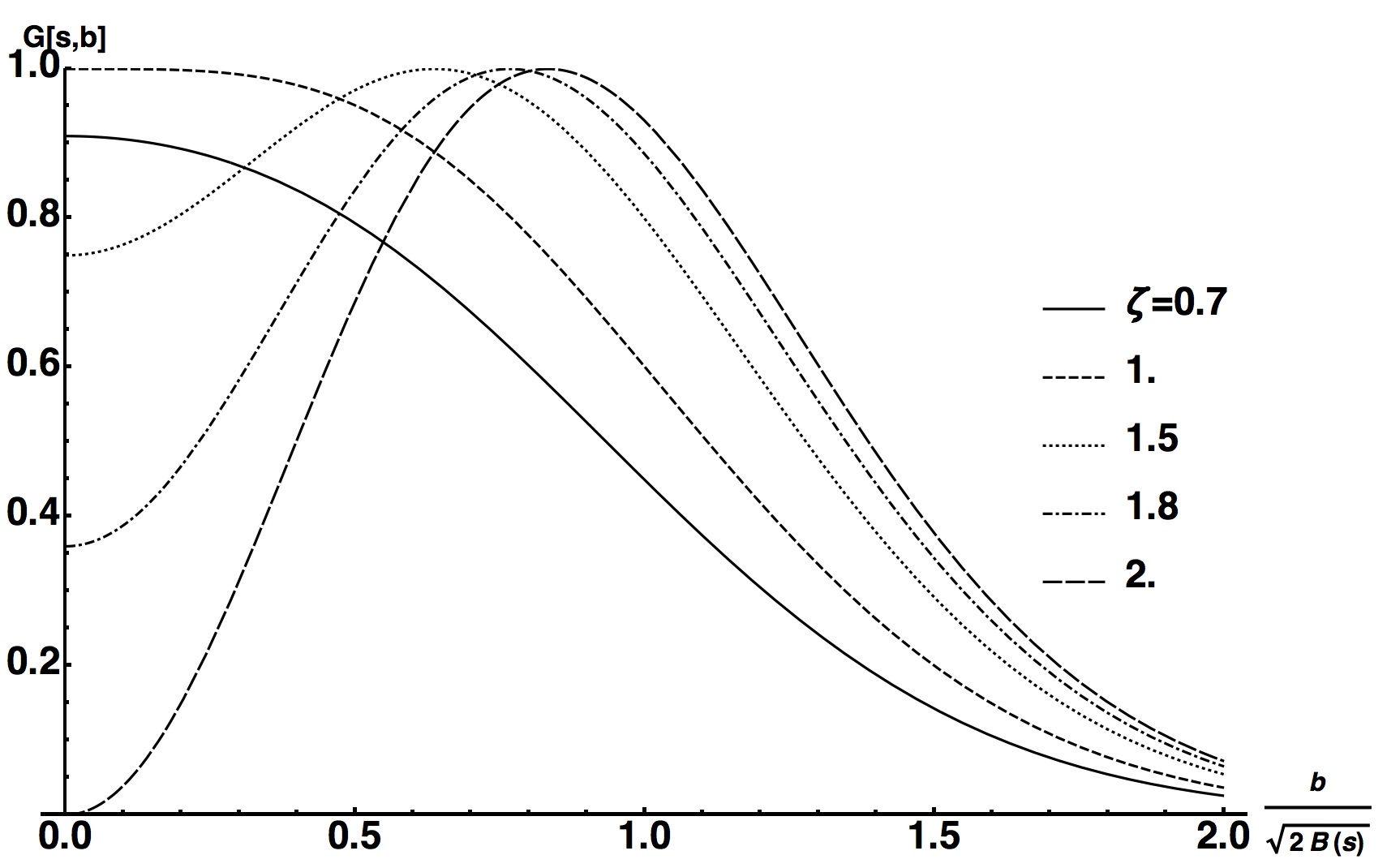}

Fig. 4. The energy evolution of the shape of the inelastic 
interaction region for different values of the survival probability
$\zeta /4$. The values $\zeta =0.7$ and $1.0$ correspond to 
ISR and LHC energies and agree well with the result of detailed fitting to the 
elastic scattering data \cite{ama, dnec, mart}. A further increase of 
$\zeta $ leads to the toroid-like shape with a dip at $b=0$.
The values $\zeta =1.5$ are proposed in \cite{kfk, fms} and $\zeta =1.8$ in
\cite{roy} as corresponding to asymptotical regimes. The value $\zeta =2$
corresponds to the "black disk" regime ($\sigma _{el}=\sigma _{in}=
0.5\sigma _{tot}$). For more discussion of the black disk and the geometrical
scaling see Refs \cite{dias, csor, fag}.
\end{center}
\end{figure}

Before discussing the predictions at higher energies, we would like 
to point out that the cross sections of inelastic processes are determined not 
only by the strength of the interaction inside the interaction region but also 
by the purely geometrical factor. 
Even though the proton interaction region is very dark at central collisions
($G(s,b)\approx 1$ inside the plateau), the cross sections of processes with 
small impact parameters $b\leq r$ are very small because the corresponding 
areas proportional to $r^2$ are small for integrals over $b\leq 0.5$ fm. 
Integrating the total and elastic terms in Eq. (\ref{ge}) up to impact 
parameters $b\leq r$ one estimates their roles for different radii $r$.
\begin{equation}
\sigma _{el}(s, b\leq r)=\sigma _{el}(s)[1-\exp (-r^2/B(s))],
\label{el}
\end{equation}
\begin{equation}
\sigma _{tot}(s, b\leq r)=\sigma _{tot}(s)[1-\exp (-r^2/2B(s))].
\label{tot}
\end{equation}
One gets that the contribution of processes at small impact parameters 
$b^2\ll 2B$ diminishes quadratically at small $r\rightarrow 0$. In particular,
inelastic processes contribute at $r\rightarrow 0$ as
\begin{equation}
\sigma _{inel}(s, b\leq r)\rightarrow \pi r^2G(s,0)+O(r^4);\;\; (r^2\ll B).
\label{in}
\end{equation}
The maximum intensity of central inelastic collisions equal 1 is at $\zeta =1$. 
The high intensity must result in high multiplicities of inelastic events.
The integral contribution of the near central region of collisions is small.
The cross sections of very high multiplicity events are also small. The
estimates show that they are quite comparable to one another. 

This property has 
been used in Ref. \cite{ads} for the explanation of jets excess observed for 
very high multiplicity events at 7 TeV compared to predictions of the 
well known Monte-Carlo models PYTHIA and HERWIG. This excess was interpreted as 
an indication on the active role of the high density gluonic component of the 
internal structure of protons at that energy. Therefrom it was concluded that 
such a component should be more properly accounted in the new versions of the 
Monte-Carlo models. That demonstrates how the knowledge about the spatial 
view of the inelastic interaction region helps in getting some conclusions
about possible omissions in the models used nowadays for description of
experimental data on jet production at LHC energise.

The spatial region of elastic scattering as derived from the subtrahend in
Eq. (\ref{ge}) is strongly peaked at low impact parameters decreasing fast at
larger values of $b$ according to the Gaussian exponent law. The contribution to 
the elastic cross section is, nevertheless, suppressed at small $b$ and comes 
mainly from impact parameters $b^2\approx 2B$. The average value of the squared 
impact parameter for elastic scattering can be estimated as
\begin{equation}
<b_{el}^2>=\sigma _{el}(s)/\pi \zeta ^2(s).
\label{b2el}
\end{equation}
Inelastic processes are much more peripheral. The ratio of the corresponding
values of squared impact parameters is
\begin{equation}
\frac {<b_{inel}^2>}{<b_{el}^2>}=\zeta \frac {8-\zeta }{4-\zeta }.
\label{in/el}
\end{equation}
This ratio exceeds 2 already at LHC energies and would become equal to 6
for (would be!) $\zeta =2$.  The peripherality of inelastic processes
compared to elastic ones increases with increase of the proportion of elastic 
collisions. Elastic collisions are more effective at the most central
interactions.

\section{Some predictions at higher energies}

What can we expect at higher energies?

The only guesses can be obtained from the extrapolation of experimental results 
at present energies to new regimes even though our previous experience teaches 
us how indefinite and even erroneous they can be as it often happened.
Nevertheless, let us try to use some assumptions relying on the fact that
we have used only such most reliable methods of getting the necessary 
information as the unitarity condition and the quite precise experimental 
data on the elastic scattering.

First, one may assume that the share of elastic scattering $\zeta $ will 
increase but approach 1 asymptotically without crossing it. In principle, such 
an assumption can be valid because the present accuracy of experimental data at 
7 and 8 TeV is not high enough and allows it. That would 
imply that its precise value at these energies is still slightly lower than 1 
within the present experimental errors. This is the only possibility 
to keep the present status of the shape of the interaction region (BEL) when the 
inelastic profile stays quite stable with slow approach to the complete
blackness at central collisions and steady increase of its range with
asymptotical saturation. That is
a kind of "the black tube" if one implies rather long longitudinal distances
as it is commonly believed for the picture with soft wee partons.

Surely, it is not excluded that the share of elastic scattering will suddenly
decrease again. Then we would come to the picture which we dealt with, say,
at ISR energies and nothing interesting happens. This possibility looks
however quite improbable. In both cases one deals with the same branch of
the unitarity condition.

Another, more interesting and intriguing possibility is further increase of the 
share of elastic processes with increasing energy. One has to consider the values 
$\zeta >1$. The transition to another branch of the unitarity condition takes
place. The BEL-scenarium described above becomes drastically changed. 
The maximum absorption appears at non-zero impact parameters. It shifts to 
positive values of impact parameters (\ref{bm}) for $\zeta >1$. Then the 
inelastic interaction region inevitably acquires the toroid-like shape TEH with 
a dip at the very center $b=0$. Most probably, if the accuracy of experimental 
data is high enough, one will observe at 13 TeV the increase of $\zeta $ above 
1 at approximately the same rate as it happened in the range of ISR to LHC where
it changed from 0.67 to 1.0 (with intermediate values of 0.8 at S$p\bar p$S
at 546 GeV and 0.9 at Tevatron at 1.8 TeV if the proton-antiproton data
are included). Then the darkness at the very central collisions $G(s,b=0)$ 
diminishes with increase of $\zeta $. The center becomes more transparent. The 
dip at the center of the interaction region with a minimum at $b=0$ should
appear instead of the flat plateau. "The black plateau" 
described at 7 TeV transforms to the toroid-like structure with somewhat lower 
darkness at the center and maximum blackness equal 1 at more peripheral 
impact parameter $b_m$ (see \cite{jetp14, drem2, ann, anis}). As it follows 
from the above formulae, this dependence is very slow near $\zeta =1$ so that 
the darkness at the center would only become smaller ,e.g., by 6$\%$ if 
$\zeta $ increases to 1.2. Therefore one can hardly expect the immediate 
drastic changes with increase of LHC energies to 13 TeV. 
Nevertheless, the forthcoming TOTEM+CMS results on elastic scattering at 13 TeV 
can be very conclusive about the general trend if the precise values of the 
diffraction cone slope $B$ and the total cross section $\sigma _{tot}$ 
(or equivalently, of the proportion of elastic processes) become 
available and the corresponding value of $\zeta $ happens to be above 1.

 The central dip becomes even deeper at larger $\zeta $. The limiting value 
$\zeta =2$ leads to complete dominance of elastic scattering at the center 
$b=0$ with $\zeta ^2=4$. It coincides with the total profile $2\zeta =4$ there.
No inelastic absorpion can be observed at the center $G(s,0)=0$.  The maximum
absorption is shifted to $b_m=\sqrt {2B\ln 2}$. Such situation can be only 
reached if the positive sign branch of the unitarity condition is applicable.

All these features are demonstrated in Fig. 4 borrowed from Ref. \cite{igse}.
Beside the demonstration of the present energy results at $\zeta =0.7$ and 
1.0 and of the limiting plot of the attenuation at $\zeta =2$ some intermediate
values 1.5 and 1.8 are shown. These values illustrate the regimes 
with further increase and asymptotical saturation of the share of elastic 
scattering.

Such regimes are predicted by some phenomenological models \cite{kfk, fms} 
which favor the situation of the increasing proportion of elastic scattering, i.e.
of $\zeta $ becoming steadily larger than 1 at higher energies. 
They are based on good fits of a large set of experimental data at present 
energies and provide some extrapolations to ever higher energies. 
The realistic estimates of their predictions at the energies 13 TeV 
and 100 TeV \cite{ksf} show that extremely high accuracy of elastic scattering 
experiments will be 
necessary to observe some effects. Both models predict that $\zeta $ will be 
only 3-4$\% $ higher at 13 TeV than that at 7 TeV. In accordance with the above 
formulae, the darkness decrease at the center of the inelastic interaction 
region is quadratically small compared to the change of $\zeta $ itself and 
becomes noticeable at the third digit only. That asks for very high precision 
of forthcoming TOTEM+CMS results at 13 
TeV. At the newly planned 100 TeV collider the value of $\zeta $ can increase 
by 13-20$\% $ from 1. It would imply 3-4$\% $ lower value of $G(b=0)$. The 
maximum blackness 1 will be reached at the impact parameters about 0.5 fm. 
The formation of the toroid-like structure proceeds very slowly with energy. 
No model predicts the fast rise of $\zeta $ to values close to 2. The 
asymptotical values of $\zeta $ preferred by both models are about 1.5. The 
corresponding asymptotical profiles of inelastic processes are shown in Fig. 4. 
The somewhat different asymptopia for 
$\zeta $ equal 1.8 is favored in the theoretical paper \cite{roy}. 
Its prediction of the deeper dip is also demonstrated in Fig. 4.
The whole impact-parameter structure in all these models reminds the toroid 
(tube) with absorbing black edges which looks as if being more and more
transparent for the elastic component at the very center. The inelastic cross
section will be only about 1.5 times larger than the elastic cross section
at asymptotics for these models. It is most fascinating in the presented 
scenarium that the density of central inelastic interactions tends to 0 
for $\zeta \rightarrow 2$ which would lead to the "black disk" limit with equal 
elastic and inelastic cross sections. However, no models predict such a high 
increase of the share of elastic scattering even at asymptotically high energies.

What concerns the inelastic processes, these models do not predict any drastic 
evolution of the interaction region with increasing energy over the LHC range. 
The (almost) black plateau with small dip at the central part near $b=0$ will 
become somewhat enlarged in size. Therefore the jet cross sections due to 
central collisions will slightly increase as well at the beginning. 
Step by step the inelastic profile will become even more peripheral
and the role of peripheral collisions will increase.  

As was discussed, central collisions are responsible for the rare events with 
highest multiplicities. The decrease of their intensity at ever higher energies
would result in lower tail of the multiplicity distributions and in their
more steepened shape. In particular, one would also predict the diminished role 
of jet production from central collisions with further increase of $\zeta $. 
Once again, these effects will develop very slowly, unfortunately.

\section{Discussion and conclusions}

The intriguing purely experimental phenomenon of the increase of the share of 
elastic processes to the total outcome observed in proton interactions at 
energies from ISR to LHC attracts much attention nowadays. It has not been 
explained yet. One of the possibilities can be related to the fact that
the larger number of the high energy constituents (quarks, gluons) exchange 
by high momenta. Due to the QCD property of the asymptotic freedom the role
of such processes would decrease, and, correspondingly, the relative role
of elastic scattering increases. Let us note that the mutual influence of the
smaller number of these processes and larger transferred momenta must lead
to some increase of the transverse momenta of created particles as observed
in experiment. Another possibility is connected to the fluctuations of the
partonic picture of colliding protons. The time of flight of protons through
one another becomes shorter with increasing energy. The pointlike partons
have almost no chance to interact during such a short time\footnote{The
classical analogy of this effect to the bullet passing through a glass was 
pointed out to me by B.L. Altshuler.}. Therefore, the role of elastic processes 
can increase.

Inspite of absence of the explanation of the observed effect,
the increase of the proportion of elastic processes has been used in this review 
paper for getting its consequences. In particular, the important information 
about the spatial regions of proton interactions has been obtained. 
The share approach to 1/4 at LHC (or, equivalently, of $\zeta $ to 1) 
can become a critical sign of the changing character of processes of hadron 
interactions if the above tendency of increase persists. The concave central 
part of the inelastic interaction region would be formed. The inelastic 
interaction region would then look like a toroid (tube) hollowed inside and
strongly absorbing in its main body at the edges. The role of elastic
scattering in central collision becomes increasing. That is surprising and
contradicts somewhat to our everyday experience and theoretical prejudices. 
Intuitively, we would expect the steady increase of the proportion of inelastic 
processes with increasing energy as it happened up to ISR. Instead of it, we
are posed to the problem that from the formal theoretical point of view the new
tendency requires now to consider another 
branch of the unitarity condition that asks for its physics interpretation.

It is hard to believe that protons become more penetrable at higher energies
after being so dark in central collisions with $G(s,0)=1$ at 7 TeV unless
some special coherence within the internal region develops. Moreover, it
seems somewhat mystifying why the coherence is more significant just for 
central collisions but not at other impact parameters where inelastic 
collisions become dominant. 

Several very speculative ideas come to the mind and have been proposed but not 
a single one looks satisfactory. Let us try to describe some of them
independently of how fantastic they look like.

For example, the role of the string junction in 
three-quark hadrons can become crucial. Then this effect would not be observed, 
say, in the pion-proton interactions. However we have no chances to get any 
experimental information about these processes. Moreover, the success of 
the quark-diquark models adds some sceptical attitude to this approach.
Probably, the relative strengths of the longitudinal and transverse components 
of gluon (string) fields can help to explain the new physics of 
TEH-scenarium of the "hollowed interactions" of protons.

In classical terms, the transparency at central collisions could reveal
itself at collisions of the two toruses with so different radii that one of 
them penetrates through the hole in another one at $b=0$. In the more general 
situation, those can be some stratified objects in which the empty spaces of one 
of them coincide at the collision with the dense regions in the another one.
They overlap at peripheral collisions and therefore lead to inelastic processes.
These fluctuations of the size and the structure of high energy protons seem 
very improbable.

One could also imagine that "black" protons start scattering 
backward \cite{igse} like the billiard balls for head-on
collisions. Snell's law admits such situation for equal reflective indices of
colliding bodies. However the forward and backward scattering can not be
distinguished for two equivalent colliding objects. That can only be checked 
if forward and backward scattered protons can be somehow identified in 
experiment. Then they should wear different labels. One can use the proton 
spin as such a label. In principle, experiments with oppositely polarised 
protons can resolve the problem. Unfortunately, no polarized protons are 
available now  even at LHC. Thus it is improbable that the TEH-structure 
will be observed directly. Moreover, the backward scattering would ask
all partons to get coherently large transferred momenta. The asymptotic
freedom of QGD claims that the probability of such processes must be
extremely low.

Beside the case of the two billiard balls colliding head-on, one could 
consider the hypothesis that centrally colliding protons 
at $\zeta =2$ remind solitons which "pass through one another without losing
their identity. Here we have a nonlinear physical process in which
interacting localized pulses do not scatter irreversibly" \cite{krus}. Again,
in the case of two identical colliding objects it is impossible to decide
whether they scatter forward or backward. In the case of solitons it is known 
that the non-linearity and dispersive properties (the chromopermittivity 
\cite{cher}) of a medium compete to produce such effect. Then one should 
understand the dynamics of the whole process. For its description one uses
the equations of Korteweg-de-Vries and sine-Gordon, the nonlinear
Schrodinger equation \cite{osch}, the Skyrme model \cite{mach}, instantons
\cite{4}. It is not at all clear yet how the QCD-nonlinearity and the 
properties of the quark-gluon medium could be responsible at the quantum-field 
level for these new features of proton interactions. Again, the
asymptotic freedom of QCD seems to forbid such processes.

Coherence of the parton structures inside the interaction region of 
colliding hadrons can probably lead to the observed effects. It can reveal 
itself in "squeezing" (or complete absorption) of the intermediate
created inelastic channels. That would lead to the antishadowing effect
with increasing role of the elastic channel which reminds the self-focusing
of the laser beams. At the model level of reggeon interactions, these
possibilities were considered in Refs \cite{dani, anni} with the discussion
of different variants of the absorbing disk.

Another more exotic hypothesis \cite{bj} which could be used to treat the 
hollowed internal TEH-region is the formation of cooler disoriented chiral 
condensate inside it ("baked-alaska" DCC). The signature of this squeezed 
coherent state would be some disbalance between the production of charged 
and neutral pions \cite{andr}, probably, noticed in some high energy cosmic 
ray experiments. However the cross sections for central 
collisions seem to be extremely small as discussed above. The failure to find 
such events at Fermilab is probably connected with too low energies available.
It leaves some hope for higher energies in view of discussions above. Total
internal reflection of coherent states from dark edges of the toroid can be 
blamed for enlarged elastic scattering (like transmission of laser beams 
in optical fibers).

The transition to the deconfined state of quarks and gluons in the central 
collisions could also be claimed responsible for new effects (see Ref. 
\cite{trty}). The optical analogy with the scattering of light on metallic 
surface as induced by the presence of free electrons is used. Again, it is hard 
to explain why that happens for central collisions while peripheral ones with 
impact parameters near $b_m$ are strongly inelastic.

To conclude, the problem of the increasing prportion of the elastic scattering 
of high energy protons, asking for its own solution, can be further studied 
only with the advent of experimental facilities of higher energy accelerators. 
Cosmic ray studies do not look very promising because of the relatively low 
accuracy of measurements. However the detailed analyses of the extensive
air showers, probably, can say something about "escaping" high energy protons.
Only very precise experimental 
results can lead to definite conclusions since the theoretically predicted 
energy dependence of the darkness of the interaction region discussed above 
is very mild. However the heuristic value of the foreseen results should not
be underestimated. If the tendency of the increasing prportion of elastic 
scattering processes persists, it would pose a problem of a new view 
on mechanisms of proton (hadron) high energy interactions. Then one should
invent new ways of explaining the transition to quite uncommon regime of
proton interactions with peculiar shapes of the interaction region.

\medskip 
{\bf Acknowlegments}

I am indebted to V.A. Nechitailo and to S.N. White for the collaboration
at different stages of our common work.
 
I gratefully acknowledge support by the RFBR-grant 14-02-00099 and 
the RAS-CERN program.

\end{document}